\documentclass[prl,aps,showpacs,twocolumn,floatfix,superscriptaddress]{revtex4}
\usepackage{epsfig}
\usepackage{color}
\usepackage{amsmath,amssymb}

\DeclareMathOperator{\csch}{csch}

\newcommand{\be}{\begin{equation}}
\newcommand{\ee}{\end{equation}}
\newcommand{\bea}{\begin{eqnarray}}
\newcommand{\eea}{\end{eqnarray}}
\newcommand{\ba}{\begin{array}}
\newcommand{\ea}{\end{array}}
\newcommand{\nn}{\nonumber}

\begin{document}
\title{Analytical results of the extensible freely jointed chain model}

\author{Alessandro Fiasconaro}
\email{afiascon@unizar.es}
\affiliation{Departamento de F\'{\i}sica de la Materia Condensada,  Universidad de Zaragoza, 50009 Zaragoza,  Spain}
\affiliation{Instituto de Biocomputaci\'on y F\'{\i}sica de Sistemas  Complejos, Universidad de Zaragoza, Zaragoza, Spain}

\author{Fernando Falo}
\affiliation{Departamento de F\'{\i}sica de la  Materia Condensada, Universidad de Zaragoza, 50009 Zaragoza, Spain}
\affiliation{Instituto de Biocomputaci\'on y F\'{\i}sica de Sistemas  Complejos, Universidad de Zaragoza, Zaragoza, Spain}

\date{\today}
\begin{abstract}
Based on classical statistical mechanics, we calculate analytically the length extension and the fluctuations, under a pulling force, of a polymer modelled as a freely jointed chain with extensible bonds, the latter considered as harmonic springs. We obtain an analytical formula for the partition function, and derive both the extension curve of the chain and the fluctuations as a function of the force. An independent high force approximation has been also evaluated. The analytical formulas have been validated by analysing the exactness of their fit on data obtained from Langevin simulations, and compared with the phenomenological expressions largely used in the past literature.
\end{abstract}



\maketitle

\section{Introduction}
The stretching curve of a long polymeric chain in a fluctuating environment has been the subject of many theoretical and experimental studies. The first experiment was performed by the Bustamante group by stretching a single DNA molecule applying a force by means of an optical tweezer \cite{Busta1992}. Their results showed that the extension curve \emph{vs} the applied force of a double stranded DNA (dsDNA), pulled at a small and intermediate force range, can be understood and depicted by means of the worm-like-chain model (WLC) \cite{wlc}, which  consists in a semiflexible continuous beam. In the most common models, the WLC model can be discretized as a chain of beads connected by sticks with the inclusion of an elastic bending, so obtaining a discrete WLC. This model improves the more naive freely jointed chain (FJC) model \cite{fjc}, composed by rigid sticks connected to each other that can freely rotate, i.e. that do not include any bending potential. The FJC model can effectively depict the elastic features of a flexible polymeric structure, i.e. having a negligible resistance to bend. An example is given by the single stranded DNA (ssDNA) whose characteristic elongation as a function of the stretching force can be satisfactorily described as a polymer without bending potential~\cite{Busta1992,storm2003}. To take into account the longitudinal elasticity of the polymers, the FJC -- as well as the WLC model -- needs a correction term not included in its simpler form. This correction has been introduced by Odijk \cite{odijk} in the WLC model by replacing the sticks with harmonic springs, and resulted in the addition of the phenomenological elastic contribution $f/(k l_0)$ to the statistical end-to-end distance of the chain obtained with \emph{inextensible} bonds. In the formula, $f$ is the applied force, $k$ the elastic constant, and $l_0$ the Kuhn length of the polymer (length of the sticks). In the inextensible case, the FJC end-to-end distance of the polymer in the direction of the applied force ($d_{ee}$), normalized with its contour length ($L_c=Nl_0$, with $N$ the number of segments), is easily calculated as the Langevin function $\mathcal{L}(\beta f l_0) = \coth(\beta f l_0) - \frac{1}{\beta f l_0}$. So, the \emph{extensible} FJC (EFJC) presents a normalized end-to-end distance:
\be
  \xi_N = \mathcal{L} (\beta f l_0) + \frac{f}{k l_0}
  \label{naive}
\ee
where $\beta=1/k_{\rm B}T$, with $k_{\rm B}$ the Boltzmann constant and $T$ the temperature of the system.
This expression has been recently used in fitting the experimental elasticity properties of some polymers~\cite{,storm2003,Grebikova2014,Grebikova2016Pol}.

Actually, a slightly different form was first used to fit the experimental data~\cite{Busta1996}:
\be
  \xi_M = \mathcal{L} (\beta f l_0) \left(1 + \frac{f}{k l_0} \right).
  \label{naiveL}
\ee
This expression has been, and still is, largely used to fit the data of different ssDNA and polymer chains~ \cite{Tskhovrebova1997,Rief1999,2002PRL_Bensimon,Wanga2001,Calderon2009,Frey2012,Bosco2014,Soler2016}

Equations (\ref{naive}) and (\ref{naiveL}) are attractively simple and handy, and they are, as said, reference formulas for extensible FJC polymer models. At a first sight they appear to reasonably agree with the experimental data, but looking closer, that agreement strongly depends on the value of the elastic constant of the polymer studied, which one of the parameters to fit. In fact, we will see that the data analysis performed by using Eq.~(\ref{naive}) are quite imprecise in modelling an EFJC model, and even worse fitting parameters are obtained by using Eq.~(\ref{naiveL}), especially for low values of the $k$ parameter.
      
Moreover, the two expressions lack of a satisfactory derivation from statistical mechanics principles in order to be completely justified in their use, and some works have presented a formal setting up of the FJC statistical mechanics model with a numerical solution \cite{Manca2012JCP,Manca2012JCP_Theory}.

In this paper we present an original analytical derivation of the partition function of the extensible FJC model and evaluate, among others, a closed formula for the \emph{end-to-end distance} $\xi_E$ as a function of the force. The formula reads:
 \be
  \xi_{E} = \mathcal{L}(\beta fl_0) + \frac{f}{kl_0} \left[1 + \frac{1-\mathcal{L}(\beta fl_0) \coth(\beta fl_0)}{1+\frac{f}{kl_0}\coth(\beta fl_0)} \right].
 \label{xi_E_1}
 \ee

An expression only valid at high forces has also been deduced by means of a complementary derivation.

Equation~(\ref{xi_E_1}) has been also derived in a different way in a previous work~\cite{2009Balavaev}, and used in a recent paper \cite{2017Radiom}.

In addition to the end-to-end distance, the analytical expression of the partition function of the EFJC model, permits the evaluation of another magnitude poorly attenctioned in the literature: the fluctuations ($\sigma$) as a function of the force. This measure can be used, in principle, as a second function able to decrease the free parameters in the fitting procedure of the experimental data. It is important to note that the fit parameters in this kind of experiments are generally three: the elastic constant $k$, the Kuhn length $l_0$, and the contour length of the polymer $L_c$. The use of a second function is then a valid strategy to reduce the degree of free parameters in the data fit, so resulting in a better control of the stretching features in data analysis.

The formulas obtained have been validated with the computer simulations of the Langevin dynamics of an EFJC polymer that moves in a fluctuating environment, confirming an excellent agreement between simulations and the expressions proposed. Moreover, we performed a number of fit analysis that showed a relevant improvement in estimating the parameters from the analytical expressions with respect to those obtained form the phenomenological formulas.

\section{The model.}
The Hamiltonian of the system is then:
\be
 H = H_0+\sum_1^N -fl_i \cos(\theta_i) + \sum_1^N \frac{1}{2} k (l_i-l_0)^2 ,
\ee
with $N$ the number of links, and $l_0$ the rest length of the spring, which corresponds to the Kuhn length of the polymer. $H_0=\sum_0^N p^2/2m$ is the kinetic energy contribution. The partition function is then the sum over all the polymer configurations of $e^{-\beta H}$, specifically the spatial angles and spring length:
\bea
 Z &=&\sum_{\{\theta_i\}\{l_i\}}e^{\beta \sum_{i=1}^N fl_i \cos\theta_i -\frac{1}{2} \beta k (l_i-l_0)^2} \nn \\
 &=& \sum_{\{\theta_i\}\{l_i\}} \prod_{i=1}^N e^{\beta fl_i \cos\theta_i-\frac{1}{2} \beta k (l_i-l_0)^2}  = \nn \\
 &=&  \prod_{i=1}^N \sum_{\{\theta_i\}\{l_i\}} e^{\beta fl_i \cos\theta_i-\frac{1}{2} \beta k (l_i-l_0)^2},
\eea
where the kinetic energy contributes with a force-independent multiplicative term, here omitted because it is not influent. All the angle configurations are independent from each other, then the partition function is factorized in the $N$ equal terms of the above product. So:
\be
 Z = \left [ \sum_{\{\theta\}\{l\}} e^{\beta fl \cos\theta-\frac{1}{2} \beta k (l-l_0)^2} \right ]^N = z^N,
 \label{z1}
\ee
where $z$ is the partition function of just one segment.

Given the continuous nature of both the angle values and the spring length, the above expression can be calculated as a spatial integral with volume element $d\Omega=l^2\sin\theta \,dl d\theta d\phi$:
 \bea
 z &=&\int_{0}^{\infty} \int_{0}^{\pi} \int_{0}^{2\pi} e^{\beta fl \cos\theta} e^{-\frac{1}{2} \beta k (l-l_0)^2} l^2\sin\theta dl d\theta d\phi = \nn \\
     &=& 4\pi \int_{0}^{\infty} \frac{\sinh (\beta fl)}{\beta f}\,\, e^{-\frac{1}{2} \beta k (l-l_0)^2} l \, dl.
 \label{FJC_z1}
 \eea
With the change of variable $\beta f l = x$, the integral of Eq.~(\ref{FJC_z1}) can be rewritten as
 \be
      z= \frac{4\pi}{\beta ^3 f^3} \int_{0}^{\infty} \sinh (x)\,\, e^{-(x-x_0)^2/2\sigma^2} x \, dx.
 \label{FJC_z1_x}
 \ee
with $\sigma^2=\beta f^2/k$.
This integral can be calculated by explicitly writing down the hyperbolic sine and making use of the tabulated integral $\int_0^{\infty} x e^{-\mu x^2 - 2\nu x} dx = \frac{1}{2\mu} -\frac{\nu}{2\mu}\sqrt{\frac{\pi}{\mu}} e^{\nu^2/\mu} [1-erf(\frac{\nu}{\sqrt{\mu}})]$. Unfortunately, the presence of the error function $erf(\cdot)$ makes the formal outcome useless for practical purposes, yet this expression can be evaluated numerically \cite{Manca2012JCP}.

\subsection{Analytical derivation.}
Despite the above difficulty, the integral of Eq.~(\ref{FJC_z1_x}) can be evaluated with a different approach, by writing it as
 \be
      z= \frac{4\pi } {\beta^3 f^2} \sqrt{\frac{2\pi \beta} {k}} \int_{0}^{\infty} x \sinh (x) G(x;x_0,\sigma)  \, dx,
 \ee
where the Gaussian term $G(x;x_0,\sigma)=e^{-\frac{(x-x_0)^2}{2 \sigma^2}} /\sqrt{2\pi \sigma^2} $ can be expanded in a series of $\delta$-functions:
 \bea
 G(x;x_0,\sigma) &=&\delta(x-x_0) + \frac{\sigma^2}{2} \frac{d^2}{dx^2}\delta(x-x_0)+... = \nn \\
                    &=&\sum_{n=0}^{\infty} \frac{(\sigma^2/2)^n}{n!} \frac{d^{2n}}{dx^{2n}}\delta(x-x_0).
 \eea
The integral can then be formally written (Weierstrass transform) as
\be
      z= A(f) \int_{0}^{\infty} x \sinh (x) e^{\frac{\sigma^2}{2} \frac{d^2}{dx^2}} \delta(x-x_0)\, dx,
\ee
with $A(f) = \frac{4\pi } {\beta^3 f^2} \sqrt{\frac{2\pi \beta} {k}} $. Because of the properties of the $\delta$-function inside the integral, \emph{i.e.} $\int_{-\infty}^{\infty} f(x) \frac{d^n}{dx^n}\delta(x-x_0) \, dx =(-1)^n\frac{d^nf(x)}{dx^n}|_{x_0}$, the resulting  approximated expression is:
 \bea
  z &=& A(f)  e^{\frac{\sigma^2}{2} \frac{d^2}{dx^2}} \left[x \sinh (x) \right]_{x_0} = \nn \\
        &=& A(f) \sum_{n=0}^{\infty} \frac{(\sigma^2/2)^{n}}{n!} \frac{d^{2n}}{dx^{2n}} \left[x \sinh (x)\right]_{x_0}.
 \eea
The above expansion is a closed form for the partition function.

The general term of the above derivative is
\bea
  \frac{d^{2n}[x \sinh (x)]}{dx^{2n}} =x\sinh(x)+2n \cosh(x)
\eea
 from which, summing up all the terms, we finally obtain:
 \bea
  z_E &=& B \frac{\sinh (\beta f l_0)}{ f } e^{\frac{\beta f^2}{2k}}\left [ 1 + \frac{f}{kl_0}\coth(\beta fl_0)\right],
   \label{z_E}
 \eea
with $B =4\pi l_0\sqrt{2\pi/\beta^3 k}$.

\vspace{0.5truecm}
\emph{End-to-end distance.---}
\begin{figure}[tb]
\centering
\includegraphics[angle=-90, width=8.1cm]{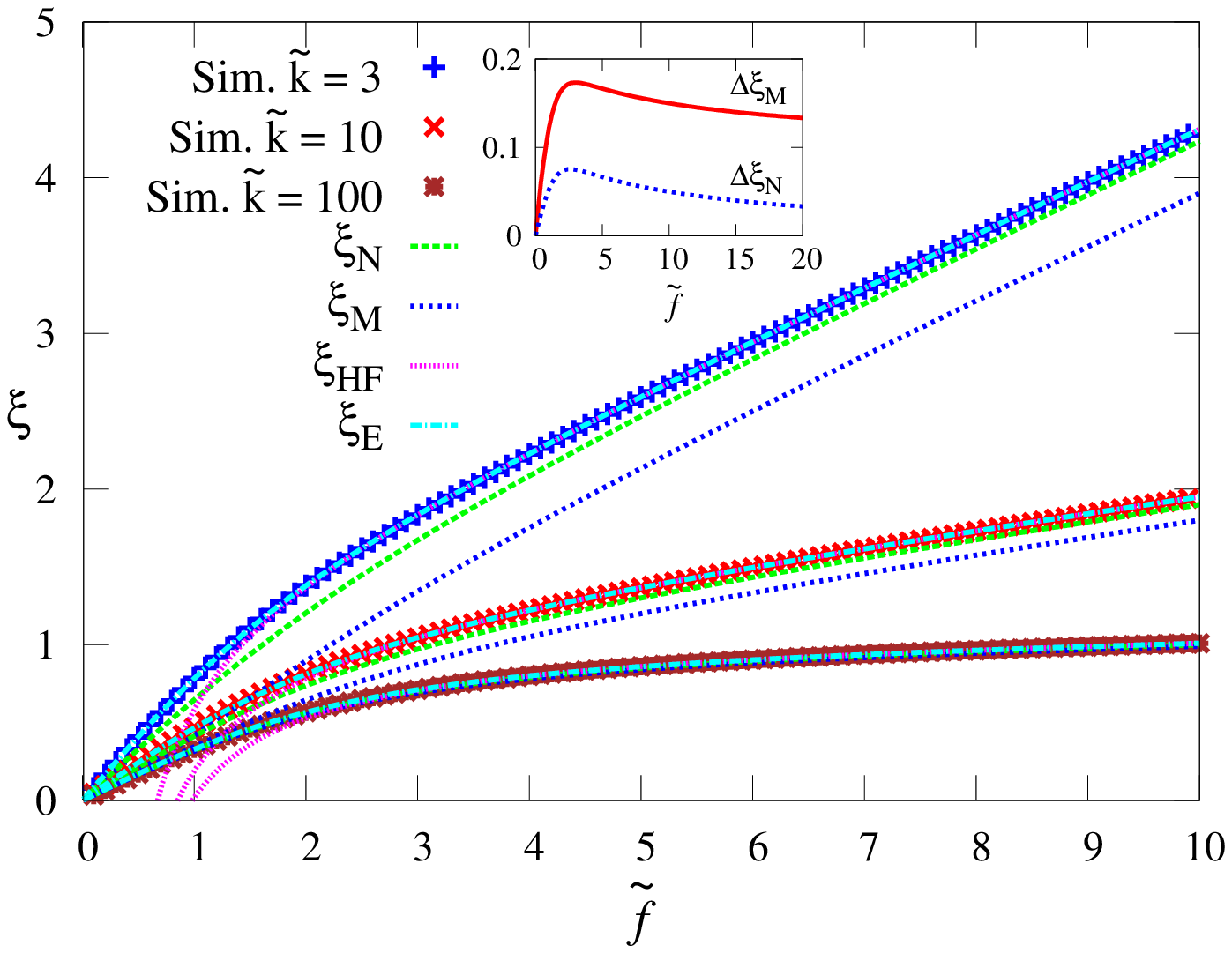}
\caption{Normalized end-to-end distance $\xi$ as a function of the dimensionless force $\tilde{f}=\beta f l_0$ for three values of the dimensionless elastic constant $\tilde{k} = \beta l_0^2 k$ in the extensible FJC model. The symbols represent the data from the simulations, and the lines the analytical expressions defined in the text. Inset: the difference between Eq.~(\ref{xi_E}) and Eq.~(\ref{naiveL}) ($\Delta\xi_M$), and between Eq.~(\ref{xi_E}) and Eq.~(\ref{naive}) ($\Delta\xi_N$), with $\tilde{k}=10$.}
\label{xi}
\end{figure}
The normalized end-to-end distance along the direction of the force is given by the average:
 \be
  \xi = \frac{1}{Nl_0} \langle l \cos\theta\rangle = -\frac{1}{Nl_0} \frac{dF}{df} = -\frac{1}{ \beta l_0 z } \frac{dz}{df}
 \label{dee}
 \ee
where $F$ is the Helmholtz free energy $F=-1/\beta \log Z$.

With this expression, the \emph{analytical expression} can be calculated by using Eq.~(\ref{z_E}), obtaining
 \be
  \xi_{E} = \mathcal{L}(\beta fl_0) + \frac{f}{kl_0} \left[1 + \frac{1-\mathcal{L}(\beta fl_0) \coth(\beta fl_0)}{1+\frac{f}{kl_0}\coth(\beta fl_0)} \right],
 \label{xi_E}
 \ee

As evident, for forces large enough, we recover the expression of $\xi_N$ given in Eq.~(\ref{naive}). Instead, none analytical limit is able to give $\xi_M$ [Eq.~(\ref{naiveL})], which remains a pure phenomenological formula.

\vspace{0.5truecm}
\emph{High force approximation.---}
For high forces, it is possible to obtain an independent approximation of the partition function. At that limit, the hyperbolic sine of the integral in Eq.~(\ref{FJC_z1}) can be substituted by the exponential with the positive exponent only:
 \be
     z = \frac{2\pi}{\beta f} \int_{0}^{\infty} e^{\beta fl} e^{-\frac{1}{2} \beta k (l-l_0)^2} l \, dl.
 \ee
With the variable change $y=l-l_0-f/k$ the integral becomes:
 \be
     z = \frac{2\pi}{\beta f} e^{\beta (fl_0 + f^2/2k)} \int_{-l_0-f/k}^{\infty} e^{-\frac{\beta k}{2} y^2} (y+l_0+f/k) \, dy.
 \label{hf}    
 \ee
As the force $f$ increases, the lower extreme of the integral shifts toward lower values, so permitting a straightforward approximation to $-\infty$ because of the sharpness of the Gaussian integrand. Then, the odd term in the integral vanishes, obtaining the simple expression:
 \bea
     z_{HF} &=& \frac{2\pi}{\beta} \sqrt{\frac{2\pi}{\beta k}} \left(l_0+\frac{f}{k}\right) \frac{e^{\beta (fl_0 + f^2/2k)}}{f}.
 \eea

By using the formula (\ref{dee}) the value of $\xi$ can be evaluated for high forces, obtaining:
 \bea
  \xi_{HF} &=& \mathcal{L}(\beta f l_0) + \frac{f}{kl_0} + \frac {1}{\beta l_0 (kl_0+f)} + 1 -\coth(\beta f l_0) \nonumber \\
  &=& 1 - \frac{1}{\beta fl_0} + \frac{f}{kl_0} + \frac {1}{\beta l_0 (kl_0+f)}.
  \label{xi_H}
 \eea

To be more clear about the validity of the above equation, it is useful to clarify what the expression ``high force" means. In this sense, the approximation applied refers to the integral of Eq.~(\ref{hf}), where the lower extreme $l_0+f/k$ goes to $\infty$. This approximation makes sense if the Gaussian inside the integral, which is centered in $0$, is narrow enough to not reach the extreme itself, i.e. 
$l_0+f/k \gg 1/\sqrt{\beta k}$
that leads to the expression $f  \gg \sqrt{k/\beta}$, so defining the relation that defines the high force regimes.

\section{Langevin simulations.}
In order to check the analytical result of equation (\ref{xi_E}) we have performed some dynamical computer simulation. In accordance with the FJC model, the polymer simulated consists of $N+1$ dimensionless monomers connected by harmonic springs:
 $ V_{\rm el}(l_i)=\frac{k}{2}\sum_{i=1}^{N} (l_i-l_0)^2,
 \label{v-har}
 $
\noindent where $k$ is the elastic constant, $l_i = |\mathbf{l}_i| = |\mathbf{r}_{i+1}-\mathbf{r}_i|$, is the distance between the monomer $i+1$ and $i$, with $\mathbf{r}_i$ the position of the $i$-th particle, and $l_0$ is the equilibrium distance between adjacent monomers.

The dynamics of the chain is given by the overdamped Langevin equation of motion
 \be
 \dot{\mathbf{r}_i} = - \mathbf{\nabla}_i V_{\rm el}(l_i) + f\delta_{i,N} + \sqrt{2k_BT} \eta(t),
 \label{eq}
 \ee
where $\eta(t)$ represents the thermal contribution as a Gaussian uncorrelated noise: $\langle\eta(t)\rangle=0$, and $\langle\eta(t)\eta(t')\rangle=\delta(t-t')$. The nabla operator is defined as $\mathbf{\nabla}_i = \partial / \partial x_i \mathbf{i} + \partial / \partial y_i \mathbf{j}  + \partial / \partial z_i \mathbf{k}$.
The constant force $f$ pulls the last monomer in order to stretch dynamically the polymer, while the first monomer is held fixed.

 Fig.~\ref{xi} shows the extension $\xi$ \emph{vs} the dimensionless applied force $\tilde{f}=\beta f l_0$, obtained from the simulations (symbols), and from the analytical formulas of Eq.~(\ref{naive}), Eq.~(\ref{naiveL}), Eq.~(\ref{xi_E}), and Eq.~(\ref{xi_H}) (lines), for different values of the elastic constant parameter  $\tilde{k}=\beta k l_0^2$. As visible in there, the numerical evaluation of the exact expression $\xi_E$ completely reproduces the simulation data at all the curve extensions, while the phenomenological formulas evidently do not. Similarly, the approximation $\xi_{HF}$ correctly approaches the curve at high forces. The figure also shows that the naive approximation $\xi_N$, while it reproduces the general behavior, lies constantly below the exact expression. The other phenomenological curve $\xi_M$, lies even lower than the previous one. These discrepancies, very well visible for $\tilde{k}=3$ and $\tilde{k}=10$, remain -- though not clearly visible in the plot -- as the elastic constant $\tilde{k}$ increases.
The inset of figure~\ref{xi} reports the differences between the exact $\xi_E$ and the two phenomenological expressions $\xi_N$ and $\xi_M$, as a function of $\tilde{f}$, for $\tilde{k}=10$. We can notice there that the difference $\Delta\xi_N$ tends to zero for $\tilde{f} \rightarrow \infty$, while $\Delta\xi_M$ tends to the value $1/\tilde{k}$. The difference between the curves is also present at low forces, where the three linear approximations read: $\xi^{LF}_N = \tilde{f}(1/3+1/\tilde{k})$, $\xi^{LF}_M = \tilde{f}/3$, and $\xi^{LF}_E = \tilde{f}[1/3+ 1/\tilde{k} + 2/(3\tilde{k}+3)]$, revealing very different slope behaviors. The expression $\xi^{LF}_E$ can be used for fit purposes by using low force data.

\begin{table}
\begin{center}
\begin{tabular}{c c c c c}
\hline \hline
Exact    & $\xi_{N}$ & $\xi_{M}$ & $\xi_{HF}$  & $\xi_{E}$  \\
$\tilde{k}$    &           &           & $f>5$       &            \\
\hline
3      &  2.86 (4.6\%)  &  2.46 (18.1\%)  &  3.00 (0.07\%)  &  3.00 (0.06\%) \\
10     &  9.18 (8.2\%)  &  7.89 (21.1\%)  & 10.00 (0.09\%)  & 10.00 (0.09\%) \\
100    &  86.7 (13.3\%) &  74.7 (25.3\%)  &  98.4 (1.63\%)  &  98.5 (1.46\%) \\
1000   &  777 (22.2\%)  &  674 (32.6\%)   &  875  (12.5\%)  &   887 (11.3\%) \\
\end{tabular}
\end{center}
\caption{Values of $\tilde{k}$ obtained by fitting the simulations data for the real parameter value listed in the first column of the table. In parenthesis, the error with respect to the exact value.}
\label{table}
\end{table}

In order to estimate the effectiveness of the formulas obtained, we have performed some numerical fit on the simulation data for different values of $\tilde{k}$, considered as the unique free parameter. The results are shown in Table~\ref{table}. As evident there, the predicted $\tilde{k}$s present an error up to $22\%$ for Eq.~(\ref{naive}), and even up to 32\% for Eq.~(\ref{naiveL}).
So, such big errors could give important discrepancies in the estimations of $k$, when the latter is used as unique fit parameter.

In principle, the complete experimental fit analysis presents up to three parameters~\cite{Busta1996,2009PRLSaleh}: the Kuhn length $l_0$, the elastic constant $\tilde{k}$, and the contour length $L_c$. This last parameter is just a multiplicative factor in all the equations (\ref{naive}), (\ref{naiveL}), and (\ref{xi_E}). In many experiments one geometrical parameter ($L_c$ or $l_0$) can be fixed from direct measures \cite{Grebikova2014,Grebikova2016Pol}, this way reducing to two the parameters to be fitted. In these conditions, the use of the formula $\xi_E$ in fitting the data provides optimum results for $\tilde{k}$ with a difference up to 30\% with the other formulas. In the case that both the geometrical magnitudes $L_c$ and $l_0$ can be independently fixed, the results in the estimation of $\tilde{k}$ are the ones shown in Table~\ref{table}. There, it is visible that the analytical formula $\xi_E$ provides estimations closer to the real values than the other expressions.

\section {Fluctuations.}
The fit of the parameters involved in the FJC model can be performed by using an independent function containing the same parameters ($l_0$, $k$, and $L_c$). In fact, it is possible to measure the end-to-end fluctuations of the chain elongation in experiments of force clamp molecular stretching by using either atomic force microscopes~\cite{2010Berkovich}, or with magnetic tweezers that present a lower intrinsic noise and relatively long constant force trajectories~\cite{2018Rafa}. The use of the fluctuations to reduce degree of freedom in data analysis has been already used to double check the parameter estimation in other contexts~\cite{Dudko2006}.

The analytic expression of the fluctuations, can be evaluated by using the second moment $\xi^{(2)}$ of the chain extension in the direction of the applied force (\emph{parallel} fluctuations) as
 \be
  \xi^{(2)} = \frac{1}{Nl_0^2} \langle l^2 \cos^2\theta\rangle =  \frac{1}{ \beta^2 l_0^2 z } \frac{d^2z}{df^2},
 \ee
where the last term in the above expression follows by the partition function of Eq.~\ref{z1}. Then, to derive the fluctuations we notice that:
 \bea
  && \frac{1}{ \beta^2 l_0^2 } \frac{d^2 \log z}{df^2}=\frac{1}{ \beta^2 l_0^2 }\frac{d}{df}\left( \frac{1}{z}\frac{dz}{df} \right) = \\ \nonumber
    &=& \frac{1}{ \beta^2 l_0^2 } \left( -\frac{1}{z^2}\frac{dz}{df}\frac{dz}{df} + \frac{1}{z}\frac{d^2 z}{df^2} \right)=\xi^{(2)} -\xi^2=\sigma^2.
 \label{fluct}
 \eea

The fluctuations $\sigma^2$ result normalized with $Nl_0^2$. By performing the above-defined derivatives to Eq.~(\ref{z_E}), we obtain the complete formula of the fluctuations $\sigma^2_{E}$:

 \begin{widetext}
 \bea
 \sigma^2_{E} =
 \frac{\csch^2(\beta f l_0)}  {2 f^2 k \beta^2 l_0^2 (k l_0 +
        f \coth(\beta f l_0))^2} \times \{
       &-& k^3 l_0^2 + f^4 \beta -
         5f^2 k^2 l_0^2 \beta + 2 f^4 k l_0^2 \beta^2 -
         2 f^2 k^3 l_0^4 \beta^2 + \\ \nonumber
       &+&
         (k^3 l_0^2 + f^4 \beta +
          f^2 k^2 l_0^2 \beta) \cosh(2 \beta f l_0) +
          2 f k l_0 (k + f^2 \beta) \sinh(2 \beta f l_0) \},
 \label{sigma}
 \eea
 \end{widetext}
that, in dimensionless magnitudes:
 \begin{widetext}
 \bea
 \sigma^2_{E} =
 \frac{\csch^2(\tilde{f})}  {2 (\tilde{k} +
        \tilde{f} \coth\tilde{f})^2} \times \{
       &-& \frac{\tilde{k}^2}{\tilde{f}^2} + \frac{\tilde{f}^2}{\tilde{k}} -
         5\tilde{k} + 2 \tilde{f}^2 - 2 \tilde{k}^2 + \\ \nonumber
       &+&
         \left( \frac{\tilde{k}^2}{\tilde{f}^2} + \frac{\tilde{f}^2}{\tilde{k}}
         + \tilde{k}\right) \cosh(2 \tilde{f} ) +
          2 \left(\frac{\tilde{k}}{\tilde{f}}+ \tilde{f} \right) \sinh(2 \tilde{f}) \}.
 \label{sigma_nodim}
 \eea
 \end{widetext}

The above expression is not very handy. An approximation can be obtained by using the truncated partition function
 \bea
   z_N &=& B \frac{\sinh (\beta f l_0)}{ f } e^{\frac{\beta f^2}{2k}},
 \label{z_E_1}
 \eea
which represent an approximation at high forces of Eq.~(\ref{z_E}), and generates formula (\ref{naive}) through Eq.~(\ref{dee}).
By using this latter function, the fluctuations corresponding to Eq.~(\ref{naive}) by means of Eq.~(\ref{fluct}) are:
 \be
  \sigma^2_{N} = \left[ 1-\coth^2(\beta f l_0)+\frac{1}{( \beta f l_0)^2} +\frac{1}{\beta k l_0^2}\right]
 \label{sigmaN}
 \ee
which, for $k \rightarrow \infty$, i.e. the inextensible case, reads
 \be
  \sigma^2_{I} = \left[ 1-\coth^2(\beta f l_0)+\frac{1}{( \beta f l_0)^2}\right]
 \ee

\begin{figure}[tbp]
\centering
\includegraphics[angle=-90, width=8.1cm]{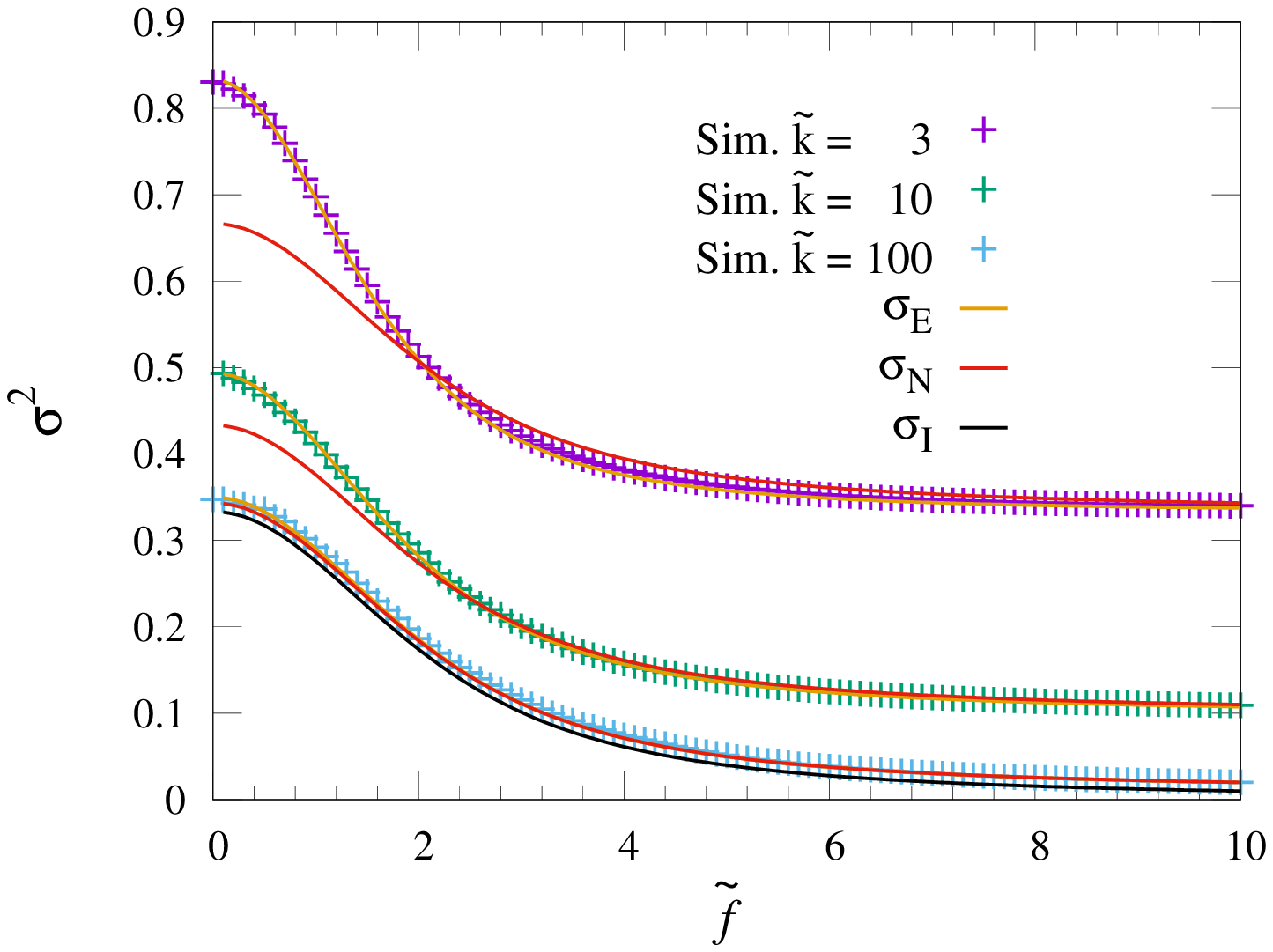}
\caption{Normalized fluctuations $\sigma^2$ as a function of the dimensionless force $\tilde{f} $ for three values of the dimensionless elastic constant $\tilde{k}$ of the extensible FJC model. The symbols represent the data from the simulations, and the lines the analytical expressions defined in the text. It is well visible that $\sigma_E$ gives good agreement with the simulations, while the phenomenological $\sigma_N$ differs visibly at low forces. The only black curve present in the figure represents the inextensible fluctuations $\sigma_I$, to which the curves tend when increasing the elastic constant value $k$.}
\label{figSigma}
\end{figure}

Fig.~\ref{figSigma} shows such fluctuations as a function of $\tilde{f}$ for different values of the elastic constant.
The curves are monotonically decreasing as a function of the applied force, and tend to the value $1/\tilde{k}$ for high forces.
The values at low forces can be easily derived from the corresponding end-to-end expressions commented above. In fact we can derive $\sigma^{2 (LF)}_{N}= 1/3+1/\tilde{k}$, and $\sigma^{2 (LF)}_{E}= 1/3+ 1/\tilde{k} + 2/(3\tilde{k}+3)$. Both these limit values are visible at low force values in Fig.~\ref{figSigma}.

Table~\ref{tableSIGMA_k} shows the evaluation of the one-parameter fit ($\tilde{k}$) by using both the fluctuations functions $\sigma_N$ and $\sigma_E$.
We can notice that, even if the value of the estimated $\tilde{k}$ is worse than the one calculated with the fit on the end-to-end distance, especially at high $\tilde{k}$ values, the fit of the fluctuations when using the expression $\sigma_E$, results always better than by using the phenomenological expression $\sigma_N$.

It is worth to note that the expressions of the fluctuations have the parameter $k$ in the denominator of some additive term. So, as higher the value of $k$ is, the smaller is its effect on the fluctuations. This explains the bad fit outcomes at high $k$s shown in Table~\ref{tableSIGMA_k}, and evidences that the fluctuations formulas are not good expressions to fit data with high $k$ values, at least when $k$ is the unique fit parameter. In fact, when the parameters to fit are more than one, the fit provides a better estimation of all of them, as visible in Table~\ref{tableSIGMA_kl}, where the two parameters $k$ and $l_0$ have been used.
\begin{table}
\begin{center}
\begin{tabular}{c c c}
\hline \hline
Exact       & $\sigma_{N}$ & $\sigma_{E}$ \\
$\tilde{k}$ &              &              \\
\hline
3           &  2.922       &    2.969     \\
10          &  9.512       &    9.917     \\
100         &  82.65       &    89.59     \\
1000        &  491.3       &    546.9     \\
\end{tabular}
\end{center}
\caption{Values of $\tilde{k}$ obtained by fitting the simulations data for the real parameter value listed in the first column of the table.}
\label{tableSIGMA_k}
\end{table}
\begin{table}
\begin{center}
\begin{tabular}{c c c c c}
\hline \hline
Exact       & $\sigma_{N}$ &     & $\sigma_{E}$ \\
$k$ &   $k$   &  $l_0$     &   $k$    & $l_0$     \\
\hline
3           &  2.62   &  1.049   &     2.96   & 0.991    \\
10          &  9.65   &  0.995   &     9.84   & 0.993    \\
100         &  105.7  &  0.974   &     88.49  & 0.999    \\
1000        &  999.6  &  0.977   &    1000.0  & 1.002    \\
\end{tabular}
\end{center}
\caption{Values of $k$ and $l_0$ obtained by fitting the simulations data for the real parameter value listed in the first column of the table by using the fluctuations formula only. The exact $l_0=1$ in all cases.}
\label{tableSIGMA_kl}
\end{table}

\section {Radius of gyration.}
Another magnitude that is straightforward to evaluate is the radius of gyration $R_G$ at $f=0$ and its dependance with the chain extensions.
By using the partition function $z$ of Eq.~(\ref{FJC_z1}), in the limit of $f \rightarrow 0$ it is easy to calculate the average length of the single bond as:
 \be
    \langle l \rangle =
 \frac{1}{z} 4\pi \int_{0}^{\infty} e^{-\frac{1}{2} \beta k (l-l_0)^2} l^3 dl,\nn
 \label{lmed_z}
 \ee
obtaining
 \be
   \langle l \rangle = l_0 \left (1+\frac{2}{1 + \beta k l_0^2} \right ).
 \label{l_med}
 \ee
Moreover, it is possible to demonstrate that for an ideal rigid chain at $f=0$~\cite{doi},
 \be
 R^2_G = \frac{\langle R^2 \rangle}{6} = \frac{N b^2}{6},
 \label{Rg}
 \ee
where $R$ is the end-to-end distance in the three-dimensional space, and $b$ is the stick length of a rigid FJC.
By substituting the average length $\langle l \rangle$ of Eq.~(\ref{l_med}) in the $b$ parameter of Eq.~(\ref{Rg}), the square radius of gyration for the extensible chain reads:
 \be
    R^2_G = \frac{N \langle l \rangle^2}{6}  = \frac{N l_0^2}{6}  \left (1+\frac{2}{1 + \beta k l_0^2} \right )^2.
 \label{Rg_ext}
 \ee
This expression has been confirmed with numerical calculations (data not shown).

\section {Global fit.}

\begin{figure}[htbp]
\centering
\includegraphics[angle=-90, width=8.1cm]{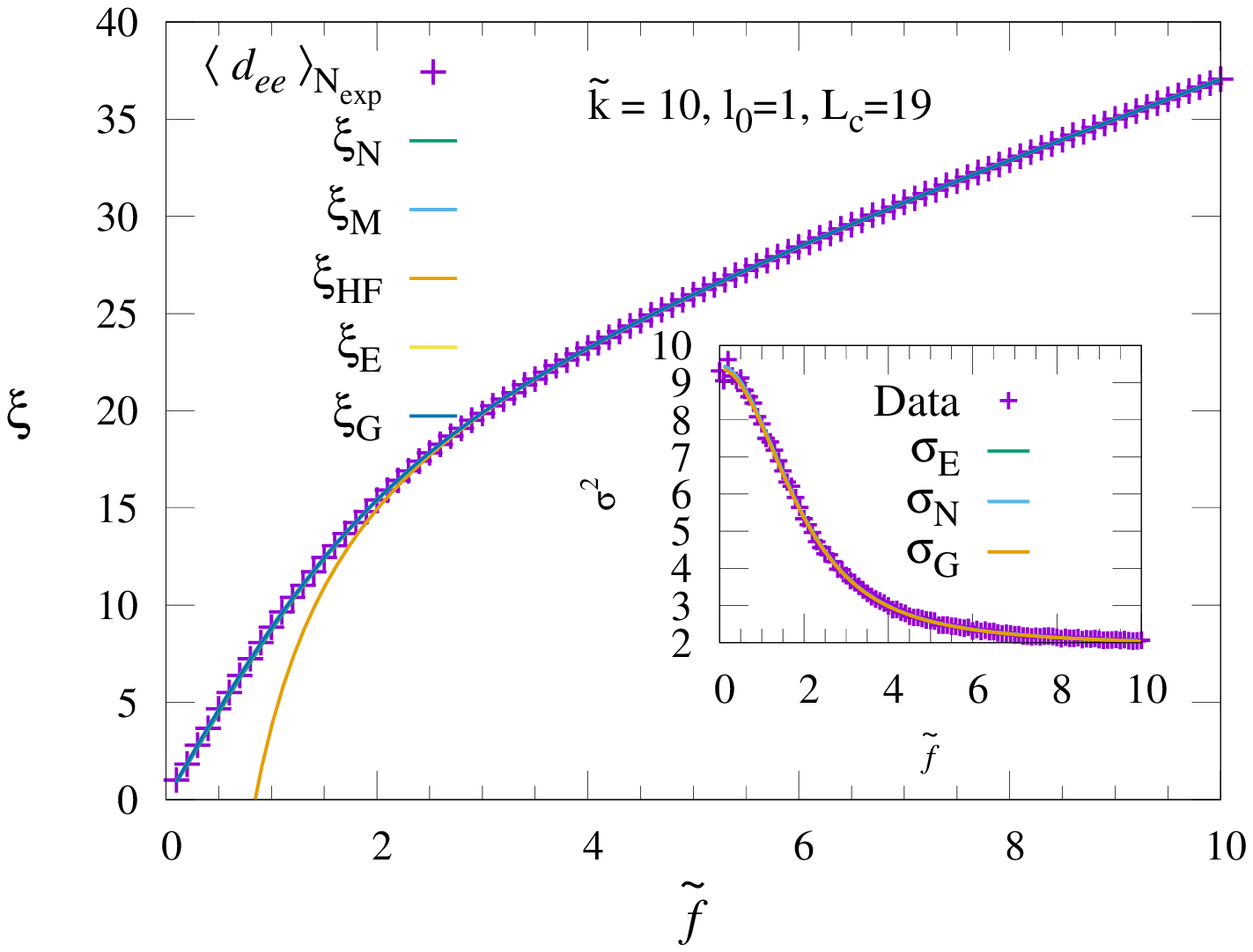}
\includegraphics[angle=-90, width=8.1cm]{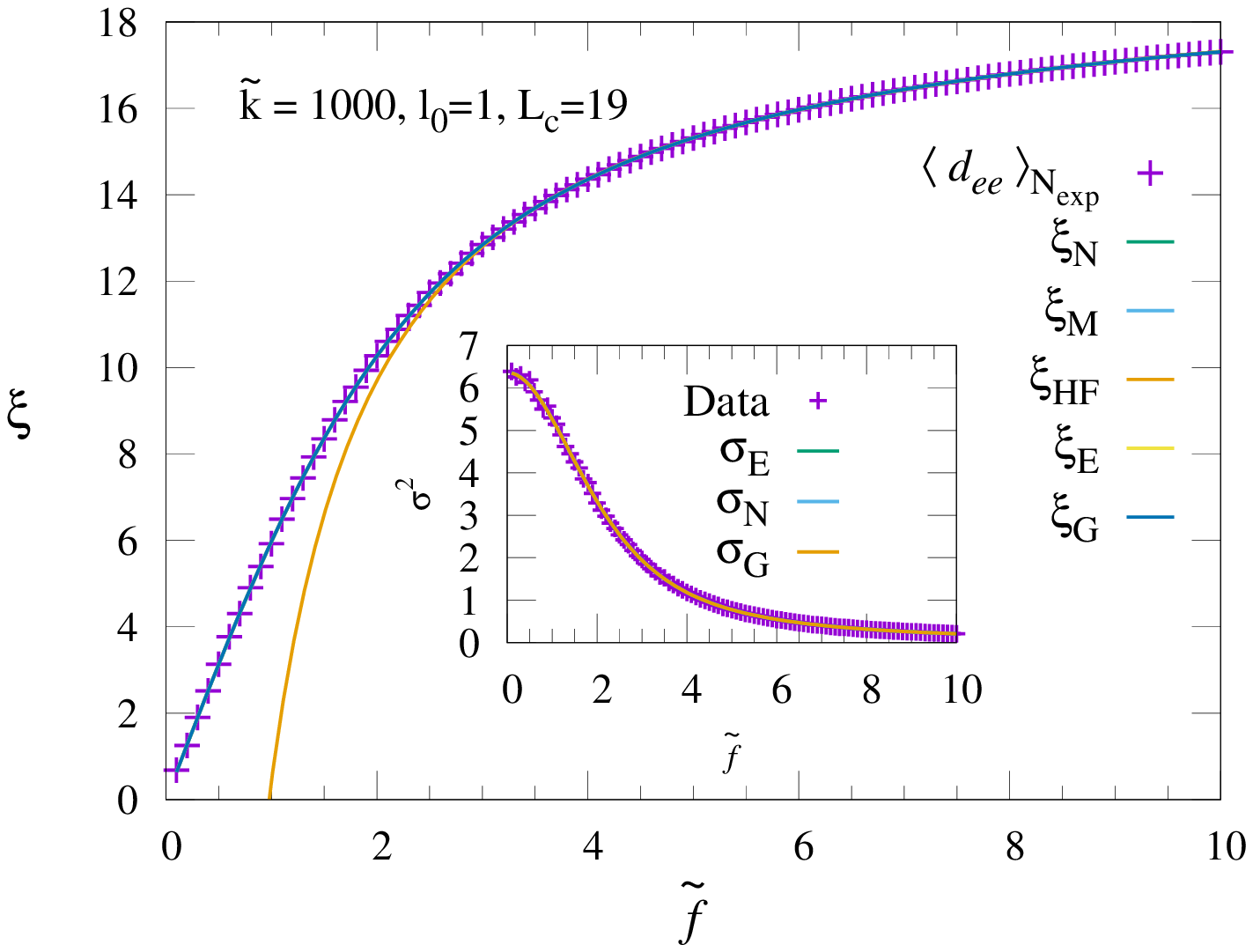}
\caption{Mean end-to-end distance $\xi$ in the FJC model as a function of the dimensionless force $\tilde{f}=\beta f l_0$ for $l_0=1$, and $L_c=19$ and two values of the elastic constant: $k = 10$ (left panel) $k = 1000$ (left panel). The symbols represent the data from the simulations ($d_{ee}$), and the lines the analytical expressions defined in the text. Insets: the simulation data and the two fits applied: the one with the Eq.~(\ref{sigma}), and the one with the global fit by using both Eq.~(\ref{xi_E}) and Eq.~(\ref{sigma}).}
\label{figXiSigma}
\end{figure}

As mentioned above, the most general case presents three free parameters. In that condition the fit analysis performed with the different end-to-end formulas give approximately the same values for the three involved parameters, due to some compensations between them. However, one can use the fluctuations as a secondary function in order to reduce the degree of freedom in the fit procedure. Fig.~\ref{figXiSigma} left panel, shows the simulation points and the fit curves for this case with $\tilde{k}=10$ and $\tilde{k}=1000$, with $l_0=1$ and $L_c=19$. The curves mostly completely overlap with the simulations, with the exception of $\xi_{HF}$ that, as expected, is only valid at high forces. The inset of the figure shows the fluctuations calculated as a function of the applied force, together with the three different fit curves, all of them overlapping: the first fit curve is the curve obtained with the fit of the simulation data of the fluctuations $\sigma_E$ with the formula (\ref{sigma}), the second curve represents the fit curve obtained by using the phenomenological formula $\sigma_N$ (Eq.~\ref{sigmaN}), and the third one shows the results from the global fit that includes the data of both the end-to-end distance and the fluctuations. The parameters obtained by the global fit have been also used to draw the respective curve in the main plot, together with the fit curves of the four end-to-end expressions above commented. Apparently, the fitting curves look fine in the plot for all the functions used.  

We also performed a fit with a more realistic value of the elastic constant $k$, specifically using $\tilde{k}=300$, and  $\tilde{k}=1000$. In our dimensionless units, the value $\tilde{k}= 300$ is equivalent to the elastic constant of a ssDNA of $530\,{\rm pN/nm}$, as estimated in~\cite{Busta1996}, while a value of $\tilde{k}= 1000$ is equivalent to the rigidity of the polimethacrylate polymer studied in~\cite{Grebikova2016Pol}, with $k=70\,{\rm nN/nm}$. Analogously, the maximum force in all our figures ($\tilde{f} = 10$) corresponds, in real units, to $f = k_{\rm B}T/l_0 \tilde{f}=4,1/l_0 \tilde{f}\,{\rm pN}$, which takes the values, respectively, of $f \approx 33\,{\rm pN}$~\cite{Busta1996} and $f \approx 170\,{\rm pN}$~\cite{Grebikova2016Pol}.

The parameters obtained in the different cases are collected in Table~\ref{tableG}, where also the end-to-end distance and the fluctuations for the inextensible model have been included, because still used in experimental works (columns $\xi_I$ and $\sigma_I$). As expected, the predicted values are very bad for low $\tilde{k}$ and tend to improve for high $\tilde{k}$s.

We can see that the outcomes of the analytical $\xi_E$ and $\sigma_E$ are very good in all the three parameters. The other outcomes, even if reasonably good for some magnitude (for example $k$), present generally worse estimations for the remaining magnitudes of the fit ($l_0$ and $L_c$). These discrepancies in the precision of the evaluation of the three magnitudes reveal the aforementioned  compensation in the fit procedure, that is evident in the fit curves which appear almost equivalent between each other in the plots. In any case, the global fit that combines together the two functions $\xi_E$ and $\sigma_E$ is generally able to even improve the already very good estimations of the two functions used separately. In fact, the three parameters involved result very close to the real ones for almost all the cases, as visible in the last column of Table~\ref{tableG} ($\xi_G$).
{Strangely enough, in many cases the analytic fluctuations only ($\sigma_E$) give better estimations than the end-to-end distance -- or not far form them -- fact that is evident at $\tilde{k}=1000$. In any case, also in this case the global fit improves, tough weakly, the outcomes for $l_0$ and $L_c$.}

We also tried the three parameters fit to the experimental data extracted from Smith et al.~\cite{Busta1996} with the three expressions evaluated, and we obtained comparable estimations between them. In this sense, the values already obtained for high k values in experiments remain reasonably good also with the new formula here presented. In order to improve and check these estimations, it would be useful to use the fluctuations data, which are not available for those known experiments.
\begin{table*}
\begin{center}
\begin{tabular}{c c c c c c c | c c c | c}
\hline \hline
   &  & $\xi_{I}$ & $\xi_{N}$ & $\xi_{M}$ & $\xi_{HF}$  & $\xi_{E}$ & $\sigma_{I}$ & $\sigma_{N}$  & $\sigma_{E}$ & $\xi_{G}$ \\
\hline
               & $k$   & --    &  9.81  &  9.86  &  9.89   &  10.11  & --    & 10.09  &  10.10  & {\bf 9.97}  \\
$\tilde{k}=10$ & $l_0$ & 0.440 &  1.204 &  1.116 &  1.008  &  0.994  & 0.571 & 1.109  &  1.003  & {\bf 1.002}  \\
               & $L_c$ & 46.14 &  25.88 &  20.36 &  18.96  &  19.05  & 41.60 & 25.41  &  19.11  & {\bf 18.99}  \\
\hline
                & $k$   & --    &  298   &{\bf 299} &  243    &  298  & --    & 311    &  310   & 302  \\
$\tilde{k}=300$ & $l_0$ & 0.938 &  1.008 & 1.004    &  1.022  & 1.001 & 0.978 & 1.001  & 0.998  & {\bf 1.000} \\
                & $L_c$ & 19.84 &  19.12 & 19.05    &  18.83  & 18.99 & 19.62 & 19.09  & 19.11  & {\bf 19.00} \\

\hline
                 & $k$   & --    &  1054  &  1055       &  605    &  1054    & --    & 1005  & {\bf 1004}  & 951  \\
$\tilde{k}=1000$ & $l_0$ & 0.980 &  1.001 & {\bf 1.000} &  1.020  &  0.999   & 0.994 & 1.001 & {\bf 1.000} & {\bf 1.000} \\
                 & $L_c$ & 19.25 &  19.05 &  19.03      &  18.83  &  19.01   & 19.19 & 19.05 & 19.01 & {\bf 19.00} \\

\end{tabular}
\end{center}
\caption{Three parameters fit by using the different analytic expressions discussed in the text. The Langevin simulations have used the reference values: $l_0=1$, and $L_c=19$, and two values of the elastic constant: $\tilde{k}=10$ and the more realistic values $\tilde{k}=300$ and $\tilde{k}=1000$. The bolded values indicate the best fit evaluation.}
\label{tableG}
\end{table*}

\section{Summary and discussion.}
This paper presents an analytical derivation from statistical mechanics principles of the partition function of the extensible FJC model from which both the mean end-to-end distance as a function of the stretching force and the fluctuations around that mean have been derived.
A formula only valid at high forces has also been calculated by means of a complementary derivation.
The expressions here derived establishes the EFJC as the most complex analytically-solvable polymer model.

A double check by means of Langevin simulations has been performed on the analytical outcomes, finding the limits of the application of a fit procedure on the formulas presented.

The formula of the end-to-end distance obtained is a combination of elementary functions simple enough to be implemented in any fit of experimental data of flexible polymers. More complicated is the expression for the fluctuations.  It is worth to note that in all the cases, the estimations given by the fit to the  fluctuations $\sigma_E$ differ very low from the exact values. In other words, the results obtained by using only the fluctuations of formula $\sigma_E$ is always a good reference expression for the estimations of the three parameters.
Moreover, we show in this paper that the simultaneous use of both formulas always reduces the number of free fit parameters, so obtaining a more feasible determination of the chain feature from the experimental data, which can improve the already good parameter estimations obtained when the two formulas are used separately.

\vspace{0.5cm}
\emph{Acknowledgments.}
This work is supported by the Spanish projects MINECO No.~FIS2017-87519-P and No.~FIS2014-55867-P, both cofinanced by Fondo Europeo de Desarrollo Regional (FEDER) funds. We also thank the support of the Arag\'on Government and Fondo Social Europeo Grant No. E19 to the FENOL group. We also want to thank Prof. L.M. Flor\'ia for useful discussions.


\end{document}